# WEIGHTED TV MINIMIZATION AND APPLICATIONS TO VORTEX DENSITY MODELS

P. ATHAVALE, R. L. JERRARD, M. NOVAGA, AND G. ORLANDI

ABSTRACT. Motivated in part by models arising from mathematical descriptions of Bose-Einstein condensation, we consider total variation minimization problems in which the total variation is weighted by a function that may degenerate near the domain boundary, and the fidelity term contains a weight that may be both degenerate and singular. We develop a general theory for a class of such problems, with special attention to the examples arising from physical models.

## 1. INTRODUCTION

Consider the weighted Total Variation functional

$$J(u) = \int_\Omega f(x)|Du| := \sup\left\{\int_\Omega u \cdot \operatorname{div}(f(x)\xi)\, dx : \xi \in C_c^1(\Omega; \mathbb{R}^n),\, |\xi| \leq 1\right\}, \quad (1.1)$$

defined for maps $u \in L^1_{loc}(\Omega)$, $\Omega \subset \mathbb{R}^n$ open bounded. We always assume that the weight $f$ is positive and continuous in $\Omega$ and possibly vanishing on some portion of $\partial\Omega$.

We are interested in the analysis of functionals of the type

$$I_\lambda(u) = J(u) + \frac{\lambda}{2}\int_\Omega |u - u_0|^2 g(x)\, dx \quad (1.2)$$

with $\lambda > 0$, $g(x) > 0$ a measurable weight function, and $u_0 \in L^2_g := L^2(\Omega, g(x)dx)$.

Notice that, when $f(x) = g(x) \equiv 1$, $I_\lambda$ corresponds to the well-known Rudin-Osher-Fatemi Total Variation based denoising model [25]. Anisotropic versions of TV functionals and applications to active contour and edge detection have been studied in [15]. Related homogeneous functionals for the description of landsliding have been also studied [20, 13]. This model can be interpreted by convex duality as a kind of non-local vectorial generalization of the classical obstacle problem. Other vector-valued generalizations can be considered, such as $J(u) = \|T \cdot Du\|$, $T$ a tensor (see for example [12] for the case $T \cdot P = \operatorname{trace}(P)$, related to a Hele-Shaw model, corresponding to $J(u) = \|\operatorname{div} u\|$), and connections with game theory (tug-of-war type games) of the dual variational problem are possible.

Due to the variety of applications of this model we are interested in a systematic study of properties of minimizers, in particular criteria for the minimizer being nontrivial, the characterization of the coincidence set, and conditions yielding the presence of flat zones and/or free boundaries.

We are particularly interested in functionals of the type of $I_\lambda$ that arise as reduced models in superconductivity and superfluidity (namely Bose-Einstein condensation) in axisymmetric 3-D domains (see e.g. [9, 10]), suited to describe the distribution of vorticity that takes place in certain asymptotic regimes. Indeed, it is proved in [9, 10] that in certain scaling limits, a Bose-Einsten condensate may be described by a velocity field $v : \Omega \to \mathbb{R}^3$, where $\Omega$ is a bounded, open subset of $\mathbb{R}^3$, that minimizes a functional of the form

$$\mathcal{E}(v) := \int_{\mathcal{D}} \rho_{TF} \left( |\nabla \times v| + \frac{1}{2} |v - \Phi|^2 \right).$$

Here

- $\rho_{TF}$ is a nonnegative integrable function known as the "Thomas-Fermi density", typically smooth in the interior of its support, corresponding to the limiting condensate density in the asymptotic regime under consideration.
- $\mathcal{D} := \{ x \in \mathbb{R}^3 : \rho_{TF}(x) > 0 \}$ is the limiting region occupied by the condensate
- $\Phi$ is a given vector field that models forcing terms.
- $\nabla \times v$ is interpreted as the vorticity. In general it is only a measure, and $\int_{\mathcal{D}} |\nabla \times v|$ must be understood as a total variation.

The model case occurs when $\rho_{TF}(x) = (\text{constant} - \text{quadratic})^+$ and $\Phi(x) = \alpha(-x_2, x_1, 0)$ for some $\alpha \in \mathbb{R}$, corresponding to a quadratic trapping potential and forcing via rotation around the $x_3$ axis. In [9] it is noted that for this $\Phi$, if $\rho$ is rotationally symmetric about the $x_3$-axis (that is, $\rho_{TF}$ has the form $\rho_{TF}(x) = \tilde{\rho}_{TF}(r, x_3)$ for $r = \sqrt{x_1^2 + x_2^2}$), then the minimizer has the form $v_* = w_*(r, x_3)(\frac{-x_2}{r^2}, \frac{x_1}{r^2}, 0)$, for $w_{min}$ minimizing the reduced energy

$$\mathcal{E}_\alpha^{red}(w) := \int_\Omega \tilde{\rho}_{TF} \left( |Dw| + \frac{(w - \alpha r^2)^2}{r} \right) dr \, dz, \tag{1.3}$$

where

$$\Omega := \{ (r, z) \in (0, \infty) \times \mathbb{R} \ : \ \tilde{\rho}_{TF}(r, z) > 0 \}.$$

Here, vortex lines are level sets of $w_*$, and "flat spots" – sets of positive measure on which $w_*$ is constant — represent regions in which the flow is irrotational. If $\Omega$ is bounded, then this can be written as a functional of the form $I_\lambda$, with coefficient functions $f = \tilde{\rho}_{TF}, g = 2\tilde{\rho}_{TF}/r$.

All our results are formulated with this application in mind. In particular, we will always assume that

$$f \in C(\Omega) \cap L^\infty(\Omega), \qquad f > 0 \text{ in } \Omega, \tag{1.4}$$

$$g \in L^\infty_{loc}(\Omega), \qquad \frac{1}{g} \in L^\infty_{loc}(\Omega), \qquad g > 0 \text{ in } \Omega. \tag{1.5}$$

Note that we do not assume that $f, g$ are bounded away from zero, or that $g$ is integrable. (We require $g \in L^1(\Omega)$ only once, in Lemma 4.2, and it will turn out that other mitigating factors make the results of this lemma available for functionals of the form (1.3), even when $g = \tilde{\rho}_{TF}/r$ is not integrable.) The second half of this paper focuses on the specific example (1.3), for which the corresponding $f, g$ described above satisfy (1.4), (1.5).

The book [1] is a good general reference for mathematical issues relating to vortices in Bose-Einstein condensates. Most prior work on 3d condensates has dealt with an asymptotic regime in which the number of vortex filaments is bounded. As proposed in [3], then justified first in [19] and in greater generality in [22], in this limit the geometry of vortex filaments is governed by a limiting functional that is closely related to what one obtains by considering in $\mathcal{E}$ above only the terms that are homogeneous of degree 1 with respect to $v$. Studies of this limiting functional are carried out in [2] and [23].

Plan of the paper: in Section 2 we revisit the theory of weighted BV functions and weighted TV functionals so as to include possibly degenerate weights, and we study general properties of minimizers such as existence, uniqueness and the Euler-Lagrange equation.

In Section 3 we investigate properties of level sets of minimizers. Our main result in this section is Proposition 3.1, where we show a uniform lower bound on the measure of superlevel sets. As a consequence, for bounded minimizers, the maximal level set is flat and has positive measure. The proof is based on a suitable weighted isoperimetric inequality which may be of independent interest.

In Section 4, under an additional regularity assumption on the weights, and using some ideas from [23], it is shown that minimizers are in fact of class $BV$ (Lemma 4.1), and in dimension $n = 2$ level sets meet orthogonally the boundary (Lemma 4.2).

In Section 5 we apply the previous theory to Bose-Einstein condensation models (for which $g \notin L^1(\Omega)$). Our main result in this section is Theorem 5.1, where we analytically derive properties of minimizers and their level sets, which represent vortex lines in the considered models. In particular, our results imply that in the limit in which condensates are described by the functional $\mathcal{E}_\alpha^{red}(\cdot)$ from (1.3), a condensate in the ground state always possesses a zone that vortx filaments do not penetrate, and in which the corresponding flow is thus irrotational.

In Section 6 we show some numerical simulations, and eventually in the appendix we prove the weighted isoperimetric inequality used in Proposition 3.1.

## 2. Weighted TV regularization – general results for rough weights

### 2.1. Preliminaries on BV functions.
The natural space for minimization of $I_\lambda$ is the weighted $L^2$ space
$$H := L_g^2(\Omega) := \{u : \Omega \to \mathbb{R} \text{ measurable} : \|u\| < \infty\}$$
with the norm and inner product
$$\|u\| := \langle u, u \rangle^{1/2}, \qquad \langle u, v \rangle := \int_\Omega g\, u\, v\, dx. \tag{2.1}$$

Note that the weight $g$ is not explicitly indicated in the notation for $\|\cdot\|$ or $\langle \cdot, \cdot \rangle$. With minimization problems in mind, we *always* work in the space $H$, even when discussing the pure weighted total variation $J(\cdot)$, without the quadratic term. Thus, all definitions that follow (subgradient, Legendre-Fenchel transform, *etc.*) are understood with respect to the $H = L_g^2$ inner product.

We also explicitly point out that, in this spirit, the domain of $J$ is always understood to be a subset of $H$:
$$Dom\, J := \{u \in H\ :\ J(u) < \infty\}. \tag{2.2}$$

Let $\Omega$ be an open subset of $\mathbb{R}^n$. A function $u \in L^1(\Omega)$ whose gradient $Du$ in the sense of distributions is a (vector-valued) Radon measure with finite total variation in $\Omega$ is called a function of bounded variation. The class of such functions will be denoted by $BV(\Omega)$. The total variation of $Du$ on $\Omega$ is defined as

$$|Du|(\Omega) = \sup\left\{\int_\Omega u \operatorname{div} z \, dx : z \in C_0^\infty(\Omega;\mathbb{R}^N), |z(x)| \leq 1 \,\forall x \in \Omega\right\}, \qquad (2.3)$$

where for a vector $v = (v_1, \ldots, v_n) \in \mathbb{R}^n$ we set $|v|^2 := \sum_{i=1}^n v_i^2$, and is denoted by $|Du|(\Omega)$ or by $\int_\Omega |Du|$. The map $u \to |Du|(\Omega)$ is $L^1_{loc}(\Omega)$-lower semicontinuous. $BV(\Omega)$ is a Banach space when endowed with the norm $\|u\| := \int_\Omega |u| \, dx + |Du|(\Omega)$.

Analogously one defines $BV_{loc}(\Omega)$ as the space of functions $u \in L^1_{loc}(\Omega)$ such that $|Du|(U) \leq +\infty$ for any $U \Subset \Omega$.

A measurable set $E \subseteq \Omega$ is said to be of finite perimeter in $\Omega$ if (2.3) is finite when $u$ is substituted with the characteristic function $\chi_E$ of $E$. The perimeter of $E$ in $\Omega$ is defined as $P(E, \Omega) := |D\chi_E|(\Omega)$. We denote by $\mathcal{L}^n$ and $\mathcal{H}^{n-1}$, respectively, the $n$-dimensional Lebesgue measure and the $(n-1)$-dimensional Hausdorff measure in $\mathbb{R}^n$.

In case $E$ is a set with finite perimeter, one can define its "reduced boundary" $\partial^* E$ as the set of points where there exists the limit

$$\lim_{\rho \to 0} \frac{D\chi_E(B_\rho(x))}{|D\chi_E(B_\rho(x))|} = \nu_E(x) \in S^{n-1}$$

and it has norm one. Then, one has the representation

$$D\chi_E = \nu_E(x)|D\chi_E| \text{ and } |D\chi_E| = \mathcal{H}^{N-1} \llcorner \partial^* E \,.$$

The first equality just follows from Besicovitch's derivation theorem for Radon measures, while the second from a careful study of the reduced boundary, see Section 3 in [7]. In particular, it follows that for any open set $A \subset \Omega$,

$$P(E, A) = \mathcal{H}^{n-1}(\partial^* E \cap A) \,.$$

If $E \subseteq \mathbb{R}^n$ is a measurable set and $x \in \mathbb{R}^n$, we define the (Lebesgue) upper density of $E$ at $x$ by

$$\overline{D}(E, x) := \limsup_{\rho \to 0} \frac{|E \cap B_\rho(x)|}{|B_\rho(x)|} \,,$$

while the density, when it exists, is simply the limit $\lim_{\rho \to 0} |E \cap B_\rho(x)|/|B_\rho(x)|$. Then it can also be shown that up to $\mathcal{H}^{n-1}$-negligible sets, the reduced boundary $\partial^* E$ of a set with finite perimeter coincides with the set of points where $E$ has density exactly $1/2$, and with the set of points where $E$ has density neither 0 nor 1.

Given $u \in BV(\Omega)$, we define

$$u^+(x) := \inf\{t : \overline{D}(\{u > t\}, x) = 0\} \quad \text{and} \quad u^-(x) := \sup\{t : \overline{D}(\{u < t\}, x) = 0\}.$$

Then, we say that $u$ is approximately continuous at $x \in \Omega$ if and only if $u^+(x) = u^-(x)$. The set of points where $u$ is not approximately continuous is called the singular set of $u$ and denoted by $S_u$. If $u$ is the characteristic $\chi_E$ of a set with finite perimeter, then $\partial^* E \subset J_{\chi_E}$ and $\mathcal{H}^{n-1}(J_{\chi_E} \setminus \partial^* E) = 0$. On the other hand, almost any level $\{u > t\}$ of a $BV$ function has finite perimeter, and there holds the co-area formula

$$|Du|(\Omega) = \int_{-\infty}^{+\infty} P(\{u > t\}, \Omega) \, dt \,. \qquad (2.4)$$

For a comprehensive treatment of functions of bounded variation, we refer the reader to [7].

2.2. **Basic theory for $J$ and $I_\lambda$.** In this section we assume that $\Omega$ is a bounded, open subset of $\mathbb{R}^n$, and that $f, g$ satisfy (1.4), (1.5). We collect here some basic results about $J$ and $I_\lambda$, see e.g. [15, 4, 14]. We first characterize the domain of the functional $J$: the following useful fact is a special case of the general results established in [6].

**Lemma 2.1.** *If $u \in L^1(\Omega)$, then $J(u) < \infty$ if and only if*

$$u \in BV_{loc}(\Omega), \quad \text{and} \quad f \text{ is } |Du|\text{-integrable,}$$

*where $|Du|$ denotes the total variation measure associated to the gradient measure $Du$. Moreover, $J(u)$ equals the integral of $f$ with respect to $|Du|$.*

We now recall a version of the coarea formula. The statement we give (in particular our convention for defining the sub/super levelsets $E_t$) is more or less forced on us by the fact that if $w \in H$ and $g \notin L^1(\Omega)$, then $\chi_{\{w>t\}} \notin H$ for $t < 0$.

**Lemma 2.2** (Coarea Formula). *Assume that $w \in \text{Dom } J$, and for $t \in \mathbb{R}$ define*

$$E_t = E_t(w) := \{x \in \Omega : \text{sign}(t)(w(x) - t) > 0\}. \tag{2.5}$$

*Then $E_t \in \text{Dom } J$ for a.e. $t$, and*

$$J(w) = \int_\Omega f|Dw| = \int_{-\infty}^\infty \left( \int_{\partial^* E_t} f d\mathcal{H}^{n-1} \right) dt = \int_{-\infty}^\infty J(\chi_{E_t}) dt.$$

*Proof.* Since $|w| > |t|$ in $E_t$, it is clear that $\chi_{E_t}^2 \leq t^{-2} w^2$, and hence that $\chi_{E_t} \in H$ for every nonzero $t$. Next, Lemma 2.1 implies that $w \in BV_{loc}$, and then the standard coarea formula implies that the set $\{x \in \Omega : w(x) > t\}$ has locally finite perimeter for a.e. $t$, and that

$$\int_\Omega f|Dw| = \int_{-\infty}^\infty \left( \int_{\Omega \cap \partial^* \{w>t\}} f d\mathcal{H}^{n-1} \right) dt.$$

Then the thesis follows from the facts that $\partial^*\{w > t\} \cap \Omega = \partial^*\{w \leq t\} \cap \Omega$ for every $t$, and that $\partial^*\{w \leq t\} \cap \Omega = \partial^*\{w > t\} \cap \Omega$ for every $t$ such that the level set $\{w = 0\}$ has measure zero, which holds for a.e. $t$. □

Similarly, we will sometimes use a "layer-cake decomposition" in the form

$$\int_\Omega \varphi(x) q(w(x), x) \, dx = \int_\mathbb{R} \text{sign}(t) \int_{E_t} \varphi(x) q_t(t, x) \, dx \, dt \tag{2.6}$$

where $w \in H$ and $\varphi \in L^2_{1/g}$, and $q(t, x)$ is a Lipschitz function (often $q(t,x) = t$) such that $q(0, x) = 0$. Indeed, in this case

$$\int_\Omega \varphi(x) q(w(x), x) \, dx =$$

$$\int_{\{w>0\}} \varphi(x) \int \chi_{\{0<t<w(x)\}} q_t(t,x) \, dt \, dx - \int_{\{w<0\}} \varphi(x) \int \chi_{\{0>t>w(x)\}} q_t(t,x) \, dt \, dx,$$

and then (2.6) follows by Fubini's Theorem.

Notice now that for a general convex positively 1-homogeneous function $J$ on a Hilbert space $H$ we have, for any $u \in Dom\, J$,

$$\eta \in \partial J(u) \quad \Leftrightarrow \quad J(v) - J(u) \geq \langle \eta, v - u \rangle \quad \text{for any } v \in Dom\, J, \tag{2.7}$$

where $\langle \cdot, \cdot \rangle$ denotes the inner product in $H$. Testing with $v = \lambda u$ for $\lambda > 0$, we obtain

$$\eta \in \partial J(u) \quad \Rightarrow \quad J(u) = \langle \eta, u \rangle. \tag{2.8}$$

We deduce from (2.7) the equivalent characterization

$$\eta \in \partial J(u) \Leftrightarrow J(u) = \langle \eta, u \rangle \text{ and } \langle \eta, v \rangle \leq J(v) \quad \text{for any } v \in Dom\, J. \tag{2.9}$$

In particular, the convex conjugate function $J^*$ corresponds to the indicatrix function $J^*(\eta) = 0$ if $\sup\langle \eta, v \rangle \leq J(v)$, and $J^*(\eta) = +\infty$ otherwise.

The following lemma, a small modification of results of Anzellotti [8], will be needed for a characterization of the subgradient $\partial J$.

**Lemma 2.3.** *For any $v \in Dom\, J \subset L^2_g$ and $\xi \in L^\infty(\Omega; \mathbb{R}^n)$ such that $\mathrm{div}(f\xi) \in L^2_{1/g}$, we define a distribution $(f\xi \cdot Dv)$ by*

$$\int_\Omega (f\xi \cdot Dv)\varphi := -\int_\Omega f\xi \cdot \nabla\varphi\, v - \int_\Omega \mathrm{div}(f\xi) v\varphi, \qquad \text{for } \varphi \in C_c^\infty(\Omega) \tag{2.10}$$

*Then $(f\xi \cdot Dv)$ is well-defined and satisfies*

$$\left| \int_\Omega (f\xi \cdot Dv)\varphi \right| \leq \|\varphi\|_\infty \|\xi\|_\infty \int_A f|Dv| \qquad \text{if } supp(\varphi) \subset A \text{ open}, \tag{2.11}$$

*and hence is a Radon measure with finite total variation.*

*Proof.* This is proved in [8, Theorem 1.5] if $f = g = 1$. In the present situation, our standing assumptions (1.4), (1.5) and the fact that $\mathrm{div}(f\xi) \in L^2_{1/g}$ guarantee that the integrals on the right-hand side of (2.10) are well-defined. Then (2.11) is clear if $v$ is smooth, and the general case follows, as in [8], by a density argument. This is straightforward since we only require approximation, in the interior of $\Omega$, of $v \in Dom\, J$ by smooth functions, and this is clearly possible in view of (1.4), (1.5). □

**Lemma 2.4.** *If $u \in Dom\, J$, then $\eta \in \partial J(u)$ if and only if there exists a vector field $\xi \in L^\infty(\bar\Omega; \mathbb{R}^n)$ such that $\|\xi\|_\infty \leq 1$,*

$$\int_\Omega (f\xi \cdot Dv) = \langle \eta, v \rangle = \int_\Omega g\, \eta\, v \qquad \text{for all } v \in Dom\, J, \tag{2.12}$$

*and*

$$\int_\Omega (f\xi \cdot Du) = J(u) = \int_\Omega f|Du|. \tag{2.13}$$

*In particular, $\mathrm{div}(f\xi) \in L^2_{1/g}$, so that $(f\xi \cdot Dv)$ makes sense for all $v \in Dom\, J$.*

*Moreover, if (2.12), (2.13) hold and $Du$ is an integrable vector field, then $\xi = \frac{Du}{|Du|}$ a.e. in $supp\,(Du) \cap \Omega$.*

Notice that, by taking $v \in C_c^\infty(\Omega) \subset Dom\, J$, we get that

$$(2.12) \implies \text{div}(f\xi) = -\eta g \quad \text{in } \Omega.$$

Formally, (2.12) also implies the boundary condition

$$\int_{\partial \Omega} v f \xi \cdot \nu = 0 \quad \text{for all } v \in Dom\, J \subset L_g^2.$$

One cannot immediately conclude that $f\xi \cdot \nu = 0$ on $\partial\Omega$, even formally or in a weak sense, without further assumptions on $g$, since membership in $H$ may force every $v \in Dom\, J$ to vanish to high order on sone subset of $\partial\Omega$. On the other hand, the boundary condition $f\xi \cdot \nu = 0$ trivially holds on portions of $\partial\Omega$ where $f$ vanishes.

*Proof.* Fix $\eta \in \partial J(u)$, and define the Banach space

$$\mathcal{M}_f(\Omega; \mathbb{R}^n) = \mathcal{M}_f := \{\mathbb{R}^n\text{-valued measures } \vec{\mu} \text{ on } \Omega \,:\, \|\vec{\mu}\|_{\mathcal{M}_f} := \int_\Omega f\, d|\vec{\mu}| < \infty\}.$$

If $v \in Dom\, J$, then $Dv \in \mathcal{M}_f$, and

$$|\langle \eta, v \rangle| \leq J(v) = \|Dv\|_{\mathcal{M}_f}. \tag{2.14}$$

Thus the linear functional $Dv \mapsto \langle \eta, v \rangle$ extends, by the Hahn-Banach Theorem, to a linear functional $\beta : \mathcal{M}_f \to \mathbb{R}$ satisfying

$$\beta(Dv) = \langle \eta, v \rangle = \int_\Omega g\, \eta\, v \quad \text{for } v \in Dom\, J \tag{2.15}$$

and

$$|\beta(\vec{\mu})| \leq \int_\Omega f\, d|\vec{\mu}| \quad \text{for all } \mu \in \mathcal{M}_f. \tag{2.16}$$

Let $\beta_0$ denote the restriction of $\beta$ to $L_f^1(\Omega; \mathbb{R}^n)$. From (2.16) we deduce that there exists a vector field $\xi$, with $\|\xi\|_\infty \leq 1$, such that

$$\beta_0(\psi) = \beta(\psi) = \int_\Omega f(x)\xi(x) \cdot \psi(x) \quad \text{for all } \psi \in L_f^1(\Omega; \mathbb{R}^n). \tag{2.17}$$

By combining (2.15) and (2.17), we see that if $v \in C_c^\infty(\Omega) \subset Dom\, J$, then the distribution $\text{div}(f\xi)$ satisfies

$$\int_\Omega \left(\frac{1}{\sqrt{g}}\text{div}(f\xi)\right)(\sqrt{g}v) := \int_\Omega f\xi \cdot Dv = \beta_0(Dv) \leq C\|v\|_H = \|\sqrt{g}v\|_{L^2(\Omega)}.$$

Since this holds for all $v \in C_c^\infty(\Omega)$, and since $g$ is bounded below on compact subsets of $\Omega$, it follows that $\text{div}(f\xi) \in L^2_{1/g}$. As a result, $\beta(Dv) = \int_\Omega \text{div}(f\xi) v$ for all compactly supported $v \in Dom\, J$.

Next, for $v \in Dom\, J$ and $\varphi \in C_c^\infty(\Omega)$, note that $D(v\varphi) = v\nabla\varphi + \varphi Dv$, from which it follows that $v\varphi \in Dom\, J$. Thus, by the linearity of $\beta$,

$$\beta(\varphi Dv) = \beta(D(v\varphi)) - \beta_0(v\,\nabla\varphi)$$
$$= -\int_\Omega \mathrm{div}(f\xi)\, v\varphi - \int_\Omega f\xi \cdot \nabla\varphi v$$
$$= \int_\Omega (f\xi \cdot Dv)\varphi.$$

Since $(f\xi \cdot Dv)$ is a finite measure, and in view of (2.16), this identity continues to hold for $\varphi \in C(\Omega)$. In particular, we may set $\varphi = 1$ to find that

$$\int_\Omega (f\xi \cdot Dv) = \beta(Dv) = \langle \eta, v \rangle \qquad \text{for } v \in Dom\, J.$$

Thus (2.12) holds, and (2.13) follows from (2.8).

Conversely, if (2.12) and (2.13) are satisfied, then it follows from (2.11) that

$$\langle \eta, v \rangle \le J(v) \text{ for all } v \in Dom\, J, \qquad \langle \eta, u \rangle = J(u).$$

As noted in (2.9), this implies that $\eta \in \partial J(u)$.

Finally, if $u \in W^{1,1}$ then $f\xi \cdot Du \in L^1(\Omega)$. So in this case (2.12), (2.13) imply that

$$\int_\Omega f(|Du| - Du \cdot \xi) = \int_{\mathrm{supp}(Du)} f(|Du| - Du \cdot \xi) = 0$$

which, since $\|\xi\|_\infty \le 1$, implies the final conclusion of the lemma. $\square$

A consequence of (2.13) is the following

**Lemma 2.5.** *Assume the hypotheses of Lemma 2.4, and define $E_t = E_t(u)$ as in (2.5). Then $\eta \in \partial J(u)$ if and only if $\mathrm{sign}(t)\eta \in \partial J(\chi_{E_t})$ for a.e. $t \in \mathbb{R}$.*

*Proof.* Assume that $\eta \in \partial J(u)$ and fix $\xi$ as in Lemma 2.4. Then, denoting by $\chi_{E_t}$ the characteristic function of $E_t$, Lemmas 2.2 (the coarea formula) and 2.4 imply

$$\int_\mathbb{R} J(\chi_{E_t})\, dt = J(u) = \int_\Omega (f\xi \cdot Du)\,. \tag{2.18}$$

On the other hand, fixing a sequence of compactly supported smooth functions $\varphi_k \nearrow \chi_\Omega$, we have

$$\int_\Omega (f\xi \cdot Du) = \lim_k \int_\Omega (f\xi \cdot Du)\varphi_k$$
$$= \lim_k \int_\mathbb{R} \mathrm{sign}(t) \int_\Omega (f\xi \cdot D\chi_{E_t})\varphi_k\, dt = \int_\mathbb{R} \mathrm{sign}(t) \int_\Omega (f\xi \cdot D\chi_{E_t})\, dt.$$

The second equality follows from (2.10) and (2.6), and the last equality follows from (2.11), (2.18), and the Dominated Convergence Theorem. Combining these identities yields

$$\int_\mathbb{R} dt \left( J(\chi_{E_t}) - \mathrm{sign}(t) \int_\Omega (f\xi \cdot D\chi_{E_t}) \right) = 0.$$

Since $|\xi| \le 1$, by (1.1) we have, for any $t \in \mathbb{R}$, $J(\chi_{E_t}) \ge \text{sign}(t) \int_\Omega (f\xi \cdot D\chi_{E_t})$, and hence for (2.18) to hold we must have, for a.e. $t \in \mathbb{R}$,

$$J(\chi_{E_t}) = \int_\Omega (f(\text{sign}(t)\xi) \cdot D\chi_{E_t}). \qquad (2.19)$$

In view of (2.12), (2.13), this states that $\text{sign}(t)\eta \in \partial J(\chi_{E_t})$. The proof of the converse is very similar. $\square$

2.3. **Properties of minimizers of $I_\lambda$.** We have the following

**Proposition 2.6.** *Given $f, g$ satisfying (1.4), (1.5), $\lambda > 0$ and $u_0 \in H$, there exists a unique minimizer $w \in \text{Dom } J$ of $I_\lambda$. This minimizer is characterized by*

$$\lambda(u_0 - w) \in \partial J(w). \qquad (2.20)$$

*In addition, if $u_0 \in L^\infty(\Omega)$, then*

$$\underset{\Omega}{\text{ess inf}}\, u_0 \le \underset{\Omega}{\text{ess inf}}\, w \le \underset{\Omega}{\text{ess sup}}\, w \le \underset{\Omega}{\text{ess sup}}\, u_0. \qquad (2.21)$$

*Proof.* Note that $I_\lambda(0) = \frac{\lambda}{2}\|u_0\|^2 < \infty$, so $\text{Dom } I_\lambda$ is nonempty. The functional $I_\lambda$ is a coercive, convex, weakly lower semicontinuous functional on $H$, so that the existence of a minimizer $w \in \text{Dom } J$ is obtained by direct methods in the calculus of variations (see [7]). Uniqueness is a consequence of the strict convexity of the functional $I_\lambda$.

If we temporarily write $Q(u) = \frac{1}{2}\|u - u_0\|^2$, then the Euler-Lagrange equation for the minimizer of $I_\lambda$ is

$$0 \in \partial I_\lambda(w) = \partial(J + \lambda Q)(w) = \partial J(w) + \lambda \partial Q(w),$$

where the final equality follows from standard convex analysis considerations, see for example [17], Proposition I.5.6. Upon rewriting this, we obtain the necessary and sufficient condition (2.20).

The uniform bound (2.21) follows by noticing that, for any $v \in BV(\Omega)$,

$$I_\lambda(\text{ess inf}\, u_0 \vee v \wedge \text{ess sup}\, u_0) \le I_\lambda(v).$$

$\square$

2.4. **Dual formulation.** Equation (2.20) is equivalent to

$$w \in \partial J^*(\lambda(u_0 - w)), \qquad (2.22)$$

where $J^*$, the convex conjugate to $J$, is given by

$$J^*(v) = \begin{cases} 0 & \text{if } \|v\|_* \le 1 \\ +\infty & \text{otherwise,} \end{cases} \qquad (2.23)$$

and

$$\|v\|_* = \sup\,\{\langle v, u\rangle : u \in H,\ J(u) \le 1\}. \qquad (2.24)$$

We may view the norm $\|\cdot\|_*$ related to the dual convex function $J^*$ as a weighted $W^{-1,\infty}$ norm. Very much as in the poof of Lemma 2.4, the Hahn-Banach theorem implies the variational characterization

$$\|v\|_* = \inf\{\|\xi\|_\infty : \int_\Omega (f\xi \cdot Dv) = \langle \eta, v\rangle \quad \text{for all } v \in \text{Dom } J\}. \qquad (2.25)$$

The following variant of a standard fact illustrates the impact of the (non)integrability of $g$.

**Lemma 2.7.** *Assume that $u_0 \in H = L^2_g$, let $w$ be the minimizer of $I_\lambda$, and define*

$$a = a_* := \begin{cases} \int g(x)u_0(x)dx / \int g(x)dx & \text{if } g \in L^1(\Omega) \\ 0 & \text{if } g \notin L^1(\Omega). \end{cases} \tag{2.26}$$

*Then*

$$w \text{ is constant} \iff w = a_* \iff \lambda \leq \lambda_c = \frac{1}{\|u_0 - a_*\|_*}. \tag{2.27}$$

*Moreover, if $g \in L^1(\Omega)$, then*

$$\int g(u_0 - w) = 0. \tag{2.28}$$

The minimizer $w$ does not need to satisfy (2.28) if $g \notin L^1$ (when the integral makes sense, which is not guaranteed). This can be seen from the concrete problems considered in Section 5, in which $u_0$ is positive and belongs to $L^1_g$, but $\|u_0\|_* < \infty$, so that $w = 0$ is the minimizer for sufficiently small values of $\lambda$.

If $n \geq 3$, there exist $u_0 \in H$ such that $\|u_0 - a_*\|_* = +\infty$, and hence $w$ is nonconstant for all $\lambda > 0$. In the unweighted case, this is essentially equivalent to the statement that $L^2$ does not embed in $W^{-1,\infty}$ for $n \geq 3$, and the presence of weights satisfying (1.4), (1.5) does not change things. Whether or not such $u_0$ exist in 2-dimensional domains depends on the behavior of $f$ and $g$ near $\partial \Omega$.

*Proof.* It follows from (2.22) and (2.23) that

$$\lambda \|u_0 - w\|_* \leq 1, \tag{2.29}$$

since $\partial J^*(\lambda(u_0 - w)) = \emptyset$ if $J^*(\lambda(u_0 - w)) = +\infty$.

If $g$ is integrable, then every constant function $c$ belongs to $L^2_g$. Then (2.29) and the definition (2.24) of the $\|\cdot\|_*$ norm imply that $\langle u_0 - w, c\rangle \leq \|u\|_*$ for every $c$, and hence that $\langle u_0 - w, 1\rangle = 0$, which is (2.28). It then follows from (2.28) that if $w$ is constant, then the constant must equal $a_*$.

On the other hand, if $g \notin L^1$, then no nonzero constant belongs to $H$, and so the only possible constant minimizer is $a_* = 0$.

In either case, the definition of $\lambda_c$ and the characterization (2.23) of $J^*$ imply that if $\lambda > \lambda_c$ then $J^*(\lambda(u_0 - a_*)) = +\infty$. Then $\partial J^*(\lambda(u_0 - a_*)) = \emptyset$, and $w = a_*$ does not satisfy (2.22).

On the other hand, if $\lambda \leq \lambda_c$ then $J^*(\lambda(u_0 - a_*)) = 0$, and it follows that $a_* \in \partial J^*(\lambda(u_0 - a_*))$. This is clear if $a_* = 0$, since $J^*$ attains its global minimum at $\lambda(u_0 - a_*)$. If $a_* \neq 0$, then necessarily $g \in L^1(\Omega)$, and then one can check that every constant function belongs to $\partial J^*(\lambda(u - a_*))$. Indeed, for any $v \in H$ and any $a \in \mathbb{R}$

$$J^*(\lambda(u_0 - w)) + \langle a, v - \lambda(u_0 - a_*)\rangle = \langle a, v\rangle \leq J^*(v).$$

□

**Remark 2.8.** Using the language of differential forms, and fixing a 1-form $\omega_0$ such that $d\omega_0 = 0$ (i.e. $\omega_0$ is closed), $(f\omega_0)_N = 0$ on $\partial\Omega \setminus K$ and $v = d^*(f \cdot \omega_0)$, we may reformulate the variational problem (2.25) as follows

$$\|v\|_* = \inf\left\{\left\|\frac{d^*\psi}{f} + \omega_0\right\|_\infty, \ \psi \in C^\infty(\Omega; \Lambda^2(\mathbb{R}^n)), \ \psi_N = 0 \text{ on } \partial\Omega \setminus K\right\}. \quad (2.30)$$

If one seeks a 2-form $\psi$ such that $(d^*\psi/f) + \omega_0$ minimizes the $L^\infty$ norm with respect to all compactly supported perturbations of $\psi$, then the resulting problem may be interpreted as a sort of (weighted) vector-valued analog of the Aronsson problem. It would be interesting to give further equivalent descriptions of the problem (2.30) in a game-theoretical framework, as the tug-of-war description of Aronsson problem (see e.g. [11], [24]).

2.5. **Level set formulation.** We shall say that a set $E \subset \Omega$ is *admissible* if $\chi_E \in Dom\, J$.
For an admissible $E \subset \Omega$, we set

$$P(E) = J(\chi_E) = \int_{\partial^* E \cap \Omega} f(x)\, d\mathcal{H}^{n-1}, \qquad A(E,t) = \text{sign}(t)\int_E (u_0 - t)g(x)\, dx.$$

and

$$I_{\lambda,t}(E) = P(E) - \lambda A(E,t)$$

Note that $A(E,t) = \text{sign}(t)\langle u_0 - t\chi_E, \chi_E\rangle$, which makes sense for an admissible set $E$.

**Lemma 2.9.** *If $w \in Dom\, J$, there holds*

$$I_\lambda(w) - I_\lambda(0) = \int_\mathbb{R} I_{\lambda,t}(E_t)\, dt.$$

*Proof.* The thesis follows from the coarea formula (Lemma 2.2) and the "layer-cake decomposition", see (2.6), applied to

$$\int_\Omega \frac{g}{2}(w - u_0)^2 - \int \frac{g}{2}u_0^2 = \int_\Omega g(x)q(w(x), x)\, dx \qquad \text{for } q(t,x) = \frac{1}{2}t^2 - tu_0(x),$$

where $E_t = E_t(w)$ is defined in (2.5). $\square$

**Proposition 2.10.** *A function $w \in Dom\, J$ minimizes $I_\lambda$ if and only if, for a.e. $t$, the set $E_t = E_t(w)$ minimizes $I_{\lambda,t}$ among all admissible $E \subset \Omega$.*

*Proof.* It is immediate from Lemma 2.9 that if $E_t$ minimizes $I_{\lambda,t}$ for a.e. $t$, then $w$ minimizes $I_\lambda$.

Conversely, notice first that from the Euler-Lagrange equation (2.20) for $I_\lambda$ and Lemma 2.5 we have $\text{sign}(t)\lambda(u_0 - w) \in \partial J(\chi_{E_t})$ for a.e. $t \in \mathbb{R}$. Hence, if $\chi_F \in Dom\, J$ we have

$$\text{sign}(t)\int_\Omega \lambda(\chi_F - \chi_{E_t})(u_0 - w)g\, dx \leq J(\chi_F) - J(\chi_{E_t}) = P(F) - P(E_t).$$

Moreover, (2.5) implies that $\text{sign}(t)(t - w)(\chi_F - \chi_{E_t}) \geq 0$, so we deduce that

$$\lambda A(F, t) - \lambda A(E_t, t) = \text{sign}(t) \int_\Omega \lambda(\chi_F - \chi_{E_t})(u_0 - t)g \, dx$$
$$\leq \text{sign}(t) \int_\Omega \lambda(\chi_F - \chi_{E_t})(u_0 - w)g \, dx \leq P(F) - P(E_t), \quad (2.31)$$

and the conclusion follows. $\square$

**Remark 2.11.** From the preceeding proof we deduce that if $F$ minimizes $I_{\lambda,t}$ then necessarily we must have equality in (2.31) and hence the minimizer $w$ satisfies

$$(t - w(x)) \cdot (\chi_F(x) - \chi_{E_t}(x)) = 0 \qquad \text{for a.e. } x \in \Omega.$$

This implies that $E_t \subset F \subset E^t := \{x : \text{sign}(t)(w(x) - t) \geq 0\}$. The sets $E_t$ and $E^t$ are called respectively the minimal and the maximal solution of the minimization problem $\min I_{\lambda,t}$. In particular, since for a.e. $t \in \mathbb{R}$ the level set $\{w(x) = t\}$ has zero measure, $E_t$ is the unique minimizer of $I_{\lambda,t}$ for a.e. $t \in \mathbb{R}$.

**Remark 2.12.** By (2.20) and (2.9) we deduce that

$$\langle \lambda(u_0 - w), \chi_F \rangle = \int_F \lambda(u_0 - w)g(x) \, dx \leq \int_{\partial^* F} f(x) \, d\mathcal{H}^{n-1} = P(F) \quad (2.32)$$

whenever $\chi_F \in \text{Dom } J$, with equality for $F = E_t$, for a.e. $t \in \mathbb{R}$, i.e.

$$\text{sign}(t) \int_{E_t} \lambda(u_0 - w)g(x) \, dx = \int_{\partial^* E_t} f(x) \, d\mathcal{H}^{n-1} = P(E_t). \quad (2.33)$$

One may wonder if for $F$ satisfying (2.33) we necessarily have that $F$ minimizes $I_{\lambda,t}$.

Since $\text{sign}(t)(t - w)\chi_{E_t} \leq 0$ it follows from (2.33) that for a.e. $t \in \mathbb{R}$,

$$\min I_{\lambda,t} = I_{\lambda,t}(E_t) = P(E_t) - \lambda A(E_t, t) = \text{sign}(t) \int_{E_t} \lambda(t - w)g(x) \, dx \leq 0. \quad (2.34)$$

Moreover, $\min I_{\lambda,t} = 0$ if and only if $t \geq \text{ess sup}_\Omega w > 0$ or $t \leq \text{ess inf}_\Omega w < 0$.

**Remark 2.13.** Since $I_{\lambda,t}(E) \leq I_{\lambda,s}(E)$ for every $E$, when $0 < t < s$, we have

$$\min I_{\lambda,t} \leq \min I_{\lambda,s} \qquad \text{for any } 0 < t < s, \quad (2.35)$$

Moreover, still assuming $s > t > 0$, we have $E_s \subset E_t$ and

$$\imath(s) - \imath(t) \stackrel{(2.34)}{=} -\int_{E_t \setminus E_s} (t - w)g(x) \, dx - \int_{E_s} (t - s)g(x) \, dx \,,$$

where $\imath(\tau) = \min I_{\lambda,\tau}$, so that

$$\lim_{s \to t^+} \frac{\imath(s) - \imath(t)}{s - t} = \int_{E_t} g(x) \, dx \,, \qquad \lim_{t \to s^-} \frac{\imath(s) - \imath(t)}{s - t} = \int_{E^s} g(x) \, dx \,,$$

where $E^s = \bigcap_{\ell < s} E_\ell = \{w \geq s\}$. Hence the map $\imath$ possesses bounded left and right derivatives at every point. Flat level sets of $w$ correspond to points of non-differentiability for $\imath$. Similar considerations apply to $t < s < 0$, with signs reversed.

## 3. Extremal level sets of the minimizer

When (2.27) is violated, the minimizer $w$ is nontrivial, and hence one may wonder whether $w$ has "flat spots" – that is, whether any level sets of $w$ have positive measure, as happens in the classical obstacle problem when the constraint is saturated or also for related models (see e.g. [13]).

The answer to this question requires more assumptions on the boundary behaviour of $f$ and $g$ with respect to those that we have made up to now. In particular, we will still assume (1.4) and (1.5), and we will also require that

$$f(x) \geq C\rho^m(x), \qquad |u_0(x)|\, g(x) \leq C\rho(x)^{m\theta}, \tag{3.1}$$

for some positive integer $m \in \mathbb{N}$ and some $\theta \in (0, 1]$ to be specified below, where

$$\rho(x) := \operatorname{dist}(x, \partial\Omega) \quad \text{for } x \in \Omega. \tag{3.2}$$

We continue to allow non-integrable $g$'s, but we require that the possible blow-up of $g$ near $\partial\Omega$ is compensated by the decay of $u_0$.

**Proposition 3.1.** *Assume that $\Omega \subset \mathbb{R}^n$ is a bounded open connected set with smooth boundary, and let $w$ minimize $I_\lambda$.*

*If $f, g, u_0$ satisfy (3.1) for some $m \in \mathbb{N}$ and $\theta \in (\frac{n+m-1}{n+m}, 1]$, then there exists some constant $a > 0$, depending on $f, g, u_0, n, \Omega, \lambda$, such that*

$$\mathcal{L}^n(E_t) \geq a \qquad \text{for every } t \text{ such that } \mathcal{L}^n(E_t) > 0. \tag{3.3}$$

*In particular, $w \in L^\infty(\Omega)$, and if either of $t^\pm := \operatorname{ess\,sup}_\Omega w^\pm$ is positive, then the corresponding level set $E^\pm := \{x \in \Omega : w(x) = \pm t^\pm\}$ has measure at least $a$.*

*Finally, if (3.1) holds with $\theta = \frac{n+m-1}{n+m}$, then $w \in L^\infty(\Omega)$.*

**Remark 3.2.** Proposition 3.1 remains valid if there exists $h : \mathbb{R}^{n-1} \to \mathbb{R}$ such that

$$\Omega = \{(x_1, x') \in \mathbb{R} \times \mathbb{R}^{n-1} \,:\, 0 < x_1 < h(x')\}, \qquad |\nabla h(x')| > 0 \text{ where } h(x') = 0,$$

and in addition $f > 0$ on $\{x \in \partial\Omega : x_1 = 0\}$. Indeed, in the proof of the proposition, the smoothness of $\partial\Omega$ is only used in invoking the weighted relative isoperimetric inequality (see Proposition 3.4 below), and this result is still valid under the hypotheses described here.

**Remark 3.3.** If $u_0 \geq 0$, so that $w \geq 0$ and $t^- = 0$, then Proposition 3.1 does not in general say whether the level set $\{w = \operatorname{ess\,inf}_\Omega w\}$ is flat.

However, if $g(x) \leq C\rho^{m\theta}$ (which in particular implies that $g$ is integrable) and $u_0 \in L^\infty$, then both extremal level sets of $w$ are always flat. Indeed, the hypotheses of the proposition are still satisfied by

$$\tilde{I}_\lambda(u) := J(u) + \frac{\lambda}{2}\int_\Omega g|u - \tilde{u}_0|^2, \qquad \text{for } \tilde{u}_0 := u_0 - c$$

for any constant $c$. Moreover, if $w$ minimizes $I_\lambda$, then $\tilde{w} := w - c$ clearly minimizes $\tilde{I}_\lambda$. By a suitable choice of $c$, one can arrange that $\tilde{u}_0 \leq 0$ in $\Omega$, and hence that $\operatorname{ess\,inf}_\Omega \tilde{w} < 0$. It follows that the level set $\{\tilde{w} = \operatorname{ess\,inf}_\Omega \tilde{w}\} = \{w = \operatorname{ess\,inf}_\Omega w\}$ is flat.

On the other hand, in the model case discussed in the next section, the results in [2] imply that $\{w = \text{ess inf}_\Omega w\}$ may have either positive or zero measure, depending on the choice of the parameters. Note that in this case, we have $g \notin L^1(\Omega)$.

The main tool in the proof of Proposition 3.1 is the following weighted relative isoperimetric inequality.

**Proposition 3.4.** *Assume that $\Omega \subset \mathbb{R}^n$ is a bounded open connected set with smooth boundary.*

*Then for every $m \in \mathbb{N}$, then there exists a constant $C = C(\Omega, m, n)$ such that for every set $E \subset \Omega$ of locally finite perimeter,*

$$\min\left(\int_E \rho^m(x)\,dx, \int_{\Omega \setminus E} \rho^m(x)\,dx\right) \leq C(\Omega, m, n)\left[\int_{\partial^* E \cap \Omega} \rho^m(x)\,d\mathcal{H}^{n-1}\right]^{\frac{n+m}{n+m-1}}. \quad (3.4)$$

The proof is deferred to the Appendix. For now, we accept the result and use it to prove that positive and negative extremal level sets of the minimizer are flat.

*Proof of Proposition 3.1.* We first claim that there exists a constant $a_1 > 0$ such that

$$\text{if } A(E,t) \leq a_1 \text{ and } 0 < \int_E \rho^m\,dx \leq \frac{1}{2}\int_\Omega \rho^m\,dx, \quad \text{then } I_{\lambda,t}(E) > 0. \quad (3.5)$$

Note that under these conditions, our hypotheses on $f$ and the relative isoperimetric inequality (3.4) imply that

$$0 < \int_E \rho^m\,dx \leq C\left(\int_{\partial^* E \cap \Omega} f\,d\mathcal{H}^{n-1}\right)^{\frac{n+m}{n+m-1}} = C P(E)^{\frac{n+m}{n+m-1}}. \quad (3.6)$$

For simplicity, we will prove (3.5) for $t > 0$; the argument for $t < 0$ is essentially the same. First, we may assume that $A(E,t) > 0$, since if $A(E,t) \leq 0$, then $I_{\lambda,t}(E) \geq P(E) > 0$ by (3.6).

Next, for an admissible $E$ with $A(E,t) > 0$, we compute

$$A(E,t) = \int_E (u_0 - t)g\,dx \leq \int_E (u_0 - t)^+ g\,dx \leq C\int_E \rho^{\theta m}\frac{(u_0 - t)^+}{u_0}\,dx$$

using our assumption (3.1) on $g$. Thus by Hölder's inequality and (3.6),

$$A(E,t) \leq C\left(\int_E \rho^m\,dx\right)^\theta \left(\int_\Omega \left(\frac{(u_0 - t)^+}{u_0}\right)^{\frac{1}{1-\theta}}\,dx\right)^{1-\theta} \leq c(t)\,P(E)^\sigma$$

for

$$\sigma = \frac{\theta(n+m)}{n+m-1} > 1, \qquad c(t) = C\left(\int_\Omega \left(\frac{(u_0 - t)^+}{u_0}\right)^{\frac{1}{1-\theta}}\,dx\right)^{1-\theta}.$$

It follows that

$$I_{\lambda,t}(E) \geq \left[\frac{A(E,t)}{c(t)}\right]^{1/\sigma} - \lambda A(E,t).$$

Since $c(t) \leq C\mathcal{L}^n(\Omega)^{1-\theta}$ for all $t$, is is easy to find $a_1 > 0$ such that $I_{\lambda,t}(E) > 0$ if $0 < A(E,t) \leq a_1$, proving (3.5).

Next, fix $a > 0$ such that if $\mathcal{L}^n(E) < a$, then the hypotheses of (3.5) are satisfied. If $0 < t < t^+$, we deduce from (2.34) that $I_{\lambda,t}(E_t) < 0$ and then from (3.5) that $\mathcal{L}^n(E_t) \geq a$, proving (3.3). It easily follows that $w \in L^\infty$ and that $\mathcal{L}^n(E^+) = \mathcal{L}^n(\cap_{0<t<t^+} E_t) \geq a$.

Finally, if $\theta = \frac{n+m-1}{n+m}$, then the above arguments show that if $A(E_t, t) > 0$, then

$$0 \geq I_{t,\lambda}(E_t) \geq A(E_t, t)\left(\frac{1}{c(t)} - \lambda\right).$$

By the Dominated Convergence Theorem, there exists some $T \geq 0$ such that $c(t) \leq \frac{1}{2\lambda}$ whenever $t \geq T$. As a result, if $t \geq T$, then $A(E_t, t) \leq 0$, which implies that $I_{\lambda,t}(E_t) \geq 0$, and hence (again by Remark 2.12) that $t \geq t^+ = \text{ess sup}_\Omega w$. This proves that $t^+ \leq T$. Since similar considerations apply to $t^- = \text{ess sup}_\Omega w^-$, it follows that $w \in L^\infty(\Omega)$. $\square$

## 4. $BV$ and boundary behavior for smooth weights

In this section we impose further decay and smoothness conditions on the weight $f$, as well as compatible decay conditions on $g$, involving the minimizer $w$ as well as the data $u_0$.

**Lemma 4.1.** *Assume that $f = \omega^m$ for some positive integer $m$ and $\omega \in C^2(\Omega)$ such that*

$$|\nabla \omega|^2 + \omega \geq c_1 > 0 \quad \text{in } \Omega. \tag{4.1}$$

*Let $w$ minimize $I_\lambda$ and assume that there exists some constant $C_1$ such that*

$$|g(w - u_0)| \leq C_1 f \quad \text{a.e. in } \Omega. \tag{4.2}$$

*Then $u \in BV(\Omega)$, and there exists some constant $C$, depending on $k, \|\omega\|_{C_2}, c_1$ and $C_1$, such that*

$$\int_\Omega |Dw| \leq CJ(w), \quad \text{and} \quad \mathcal{H}^{n-1}(\partial^* E_t \cap \Omega) \leq C \int_{\partial^* E_t \cap \Omega} f d\mathcal{H}^{n-1} \quad \text{for a.e. } t.$$

Assumption (4.2) is in general hard to verify, since it requires knowing something about the minimizer $w$. However,
- it is satisfied if $g \leq Cf$ and $u_0 \in L^\infty$.
- If $g$ does not decay at $\partial\Omega$, and in particular if $g \notin L^1(\Omega)$, it can also be proved to hold for certain $u_0$. Indeed, we prove in Section 5 that this is the case for the problems related to Bose-Einstein condensation.

The proof is an adaptation to our setting of an argument from Section 3 of [23], the main difference being simply that we use the first variation formula and an approximation argument in place of the direct ODE arguments of [23].

*Proof.* Consider $V \in C_c^1(\Omega; \mathbb{R}^n)$, and let

$$\Phi^h(x) := x + hV(x)$$

Then there exists $h_0 > 0$ such that if $|h| < h_0$, then $\Phi^h$ is a diffeomorphism of $\Omega$ onto itself which fixes a neighborhood of $\partial\Omega$.

Recall from Proposition 2.10 that $E_t = E_t(w)$ minimizes $I_{\lambda,t}(\cdot) = P(\cdot) - \lambda A(\cdot, t)$ among all admissible sets $F$, that is, sets such that $\chi_F \in Dom\, J$. Consider any $t$ for which this holds, and for $|h| < h_0$ let $E_t^h := \Phi^h(E_t)$. Since $\Phi^h$ is $C^1$, standard theory implies that $E_t^h$ is a set of locally finite perimeter in $\Omega$. Moreover, $E_t^h$ coincides with $E_t$ in a neighborhood of $\partial\Omega$, and thus is easily seen to be admissible. Then the minimality of $E_{\lambda,t}$ implies that

$$0 = \frac{d}{dh}\bigg|_{h=0} I_{\lambda,t}(E_t^h)$$
$$= \int_{\partial^* E_t}\left[f(\delta_{ij} - \nu_i\nu_j)V^i_{x_j} + V^i f_{x_i}\right] d\mathcal{H}^{n-1} - \lambda\,\mathrm{sign}(t) \int_{\partial^* E_t} g(u_0 - t) V \cdot \nu\, d\mathcal{H}^{n-1},$$

where $\nu(x)$ denotes the approximate outer unit normal vector to $E_t$ at $x \in \partial^* E_t$. Indeed, the first integral on the right-hand side is the first variation of the weighted perimeter $P(\cdot)$, see for example [21], chapter 17, and the second integral on the right-hand side is the first variation of $A(\cdot, t)$.

Let now $V = q(\omega)\nabla\omega$, for some smooth function $q$ such that

$$0 \leq q \leq 1, \qquad q' \geq 0, \qquad q = 0 \text{ in a neighborhood of } 0.$$

Since $q \circ \omega$, and hence $V$, are compactly supported, we deduce that

$$\int_{\partial_* E_t} q(\omega) m \omega^{m-1} |\nabla\omega|^2 d\mathcal{H}^{n-1} = -\int_{\partial_* E_t} \omega^m q'(\omega)(I - \nu\otimes\nu):(\nabla\omega \otimes \nabla\omega)\, d\mathcal{H}^{n-1}$$
$$- \int_{\partial_* E_t} \omega^m q(f)(I - \nu\otimes\nu):\nabla^2\omega\, d\mathcal{H}^{n-1}$$
$$- \lambda\,\mathrm{sign}(t) \int_{\partial_* E_t} q(f)g(u_0 - t)\nabla\omega \cdot \nu\, d\mathcal{H}^{n-1}.$$

For a.e. $t$, it follows from (4.2) that $|g(u_0 - t)| \leq C_1 f = C_1 \omega^m$ almost everywhere on $\partial^* E_t$. Using this and the fact that $q' \geq 0$, we deduce that

$$\int_{\partial_* E_t} q(\omega) m \omega^{m-1} |\nabla\omega|^2 d\mathcal{H}^{n-1} \leq -\int_{\partial_* E_t} \omega^m q(\omega)(I - \nu\otimes\nu):\nabla^2\omega\, d\mathcal{H}^{n-1}$$
$$- \lambda\,\mathrm{sign}(t) \int_{\partial_* E_t} q(\omega)g(u_0 - t)\nu \cdot \nabla f\, d\mathcal{H}^{n-1}$$
$$\leq \|\omega\|_{C^2}(1 + \lambda C_1) \int_{\partial_* E_t} \omega^m.$$

Letting $q \nearrow \chi_{\mathbb{R}^+}$, we conclude that

$$\int_{\partial^* E_t \cap \Omega} \omega^{m-1} |\nabla\omega|^2 d\mathcal{H}^{n-1} \leq C \int_{\partial^* E_t \cap \Omega} \omega^m\, d\mathcal{H}^{n-1}.$$

for $C = C(\|\omega\|_{C^2}, C_1, \lambda)$, for a.e. $t$. Since $|\nabla\omega|^2 + \omega \geq c_1 > 0$ in $\Omega$, it follows that

$$\int_{\partial^* E_t \cap \Omega} \omega^{m-1} d\mathcal{H}^{n-1} \leq \frac{1}{c_1}\int_{\partial^* E_t \cap \Omega} \omega^{m-1}(|\nabla\omega|^2 + \omega) d\mathcal{H}^{n-1} \leq C\int_{\partial^* E_t \cap \Omega} \omega^m.$$

Repeating essentially same argument, with $V$ replaced successively by $q(\omega)\omega^{-j}\nabla\omega$ for $j = 1, \ldots, m-1$, we find that

$$\int_{\partial^* E_t \cap \Omega} \omega^{m-j-1} d\mathcal{H}^{n-1} \leq C\int_{\partial^* E_t \cap \Omega} \omega^{m-j} d\mathcal{H}^{n-1}, \qquad 0 \leq j \leq m-1$$

for a.e. $t$. Putting together these estimates, we conclude that $\mathcal{H}^{n-1}(\partial^* E_t) \leq CP(E_t)$ for a.e. $t$. The estimate $\int_\Omega |Dw| \leq CJ(w)$ then follows from the coarea formula. □

The next result is due to Montero and Stephens (see Section 4 of [23]) in a slightly different setting. For this we specialize to $n = 2$, and we impose still stronger assumptions on $f, g, u_0$.

**Lemma 4.2.** *Assume that $n = 2$, that the hypotheses of Lemma 4.1 are satisfied, and that in addition*
$$g \in W^{1,\infty}(\Omega), \qquad u_0 \in W^{1,\infty}(\Omega).$$
*Then for a.e. $t$, $\partial E_t$ is a union of $C^2$ curves meeting orthogonally $\partial \Omega$.*

We recall the argument from [23]; this will be needed later when we argue that the result still applies to certain models of Bose-Einstein condensates for which $g \notin L^1(\Omega)$.

*Proof.* Fix some $t$ such that $E_t$ minimizes $I_{\lambda,t}$ and $\mathcal{H}^1(\partial^* E_t) < \infty$ and such that for $\beta := g(u_0 - t)$, we have $|\beta| \leq C_1 f$ almost everywhere on $\partial^* E_t$. (These conditions are satisfied a.e. by Proposition 2.10, Lemma 4.1 and (4.2).) The minimality of $E_t$, together with standard regularity results, implies that $\partial E_t \cap \Omega$ is a countable union of nonintersecting $C^2$-curves, each of which must have finite arclength. Let $\gamma : (a, b) \to \Omega$ be an arclength parametrization of one such curve. The Euler-Lagrange equation for $I_{\lambda,t}$ may be written (along $\gamma$) in the form
$$-(f(\gamma(s))\gamma'(s))' + \nabla f(\gamma(s)) - \lambda \beta(\gamma(s))\nu(s) = 0$$
where $\nu(s)$ is the outer unit normal to $\partial E_t$ at $\gamma(s)$. Suppose that $\gamma(s)$ approaches $\partial \Omega$ as $s \searrow a$ or $s \nearrow b$; we consider the former case for concreteness, and we assume that $a = 0$. Then by integrating the above equation from 0 to $s$, taking the inner product with $\nu(s)$, and rewriting, we find (since $f = 0$ on $\partial \Omega$) that
$$0 = \frac{1}{s} f(\gamma(s))\gamma'(s) \cdot \nu(s)$$
$$= \nu(s) \cdot \nabla f(\gamma(s)) + \frac{1}{s} \int_0^s \nu(s) \cdot \left[(\nabla f(\gamma(t)) - \nabla f(\gamma(s))) - \lambda \beta(\gamma(t))\nu(t)\right] dt .$$

Using this and hypotheses (4.1), (4.2) one can verify that $\nu(s) \cdot \nabla \omega(\gamma(s)) \to 0$ as $s \searrow a$, which implies that the curve meets $\partial \Omega$ orthogonally. □

## 5. Applications to quantum fluids

As mentioned in the Introduction, in certain limits the velocity field in rotationally forced Bose-Einstein condensate is described by a particular instance of the functional $I_\lambda$ from (1.2), where $\lambda = 2$ and $f, g, \Omega$ have the form
$$\Omega := \{(r, z) \in (0, \infty) \times \mathbb{R} : \rho_{TF}(r, z) > 0\}, \qquad f = \rho_{TF}, \qquad g(r, z) = \frac{1}{r}\rho_{TF}. \qquad (5.1)$$
Here $\rho_{TF} : \mathbb{R}^2 \to \mathbb{R}$ is a smooth function known as the "Thomas-Fermi density" and determined, in physical experiments, by the form of a potential that is used to confine the condensate. We will assume that

$$\rho_{TF} + |\nabla \rho_{TF}| > 0 \text{ on } \partial\Omega, \qquad \rho_{TF} \text{ is } C^2, \qquad \Omega \text{ is bounded}. \qquad (5.2)$$

The rotational forcing is encoded in the function $u_0$, which has the form

$$u_0(r, z) = \alpha r^2 \qquad (5.3)$$

where $\alpha > 0$ is a parameter.

The following results about this functional are mostly immediate consequences of the general theory developed above.

**Theorem 5.1.** *Let $I^\alpha(\cdot)$ denote $I_\lambda(\cdot)$ for $\lambda = 2$ and $f, g, u_0$ as in (5.1), (5.3), where $\rho_{TF}$ satisfies (5.2) and either*

$$\frac{\partial \rho_{TF}}{\partial r}(r, z) \leq 0 \qquad \text{in } \Omega \qquad (5.4)$$

*or*

$$\Omega \text{ is a compact subset of } (0, \infty) \times \mathbb{R}. \qquad (5.5)$$

*Then for every $\alpha > 0$, there exists a unique minimizer $w_\alpha \in \text{Dom } J$ of $I^\alpha$.*

(1) *$w_\alpha$ is the unique element of $H = L_g^2$ satisfying the equivalent conditions:*
   - *the Euler-Lagrange inclusion (2.20);*
   - *the dual formulation (2.22);*
   - *the level set formulation from Proposition 2.10.*

(2) *$w_\alpha$ is constant (corresponding to a vortex-free condensate) if and only if*

$$\alpha \leq \frac{1}{2\|r^2 - c_*\|_*} \qquad \text{for } c_* = \begin{cases} 0 & \text{if (5.4) holds} \\ \int_\Omega g r^2 dr\, dz / \int_\Omega g\, dr\, dz & \text{if (5.5) holds} \end{cases}$$

*where the $\|\cdot\|_*$ norm is defined in (2.24).*

(3) *If we write $I_t^\alpha(E) := \int_{\partial^* E} f\, d\mathcal{H}^1 - 2 \int_E g(\alpha r^2 - t)$, then*

$$\|w_\alpha\|_{L^\infty} = \inf\{t > 0 \;:\; I_t^\alpha(E) > 0 \text{ for all admissible } E \subset \Omega\}. \qquad (5.6)$$

(4) *The level set $\{x \in \Omega : w_\alpha = \max_\Omega w_\alpha\}$ has positive measure.*

(5) *If (5.5) holds, then the level set $\{x \in \Omega : w_\alpha = \inf_\Omega w_\alpha\}$ has positive measure.*

(6) *If (5.4) holds, then $0 \leq w_\alpha \leq u_0$. As a result, $w_\alpha = \frac{\partial w_\alpha}{\partial \nu} = 0$ on $\partial\Omega \cap \{r = 0\}$.*

(7) *$w_\alpha \in BV(\Omega)$, and $\frac{\partial w_\alpha}{\partial \nu} = 0$ on $\partial\Omega \cap \{r > 0\}$ in the sense that a.e. level set of $w_\alpha$ meets $\partial\Omega$ orthogonally.*

(8) *If (5.4) holds, and if $E_t^i$ denotes a connected component of $E_t$ for some $t > 0$, then $E_t^i$ is simply connected and $\partial E_t^i \cap \partial\Omega \neq \emptyset$.*

**Remark 5.2.** Both the hypotheses (5.4), (5.5) are motivated by physical experiments. Indeed, assumption (5.4) covers any convex trapping potential that is symmetric about the vertical axis. In particuar, this includes the model case of a quadratic trapping potential, corresponidng to $\rho_{TF}(r, z) = \rho_0 - \frac{r^2}{a^2} - \frac{z^2}{b^2}$ for positive constants $\rho_0, a, b$. Note that when (5.4) holds, necessarily $g \notin L^1(\Omega)$.

The second case (5.5) is relevant to toroidal Bose-Einstein condensates.

**Remark 5.3.** Recall that the level set formulation states that for a.e. $t > 0$, the level set $E_t = \{w > t\}$ of $w_\alpha$ minimizes the functional $I_t^\alpha$ appearing in conclusion (3). It follows that $\partial E_t$ is analytic in $\Omega$ and satisfies the equation

$$\operatorname{div}(\rho_{TF}\nu) = \rho_{TF}\kappa + \nabla\rho_{TF} \cdot \nu = 2\rho_{TF}\left(\alpha r - \frac{t}{r}\right) \tag{5.7}$$

where $\nu$ and $\kappa$ denote respectively the exterior unit normal and the curvature of $\partial E_t$. (This equation appears already in a different form in the proof of Lemma 4.2 above.)

*Proof of Theorem 5.1.* In case (5.5), $f$ and $g$ are smooth in $\bar{\Omega}$, and both are comparable to the distance $\rho(\cdot)$ to $\partial\Omega$, so all the hypotheses of every result in Sections 2 - 4 are satisfied (with $m = 1$ for Section 3), and all the conclusions above follow, after noting that conclusion (3) is a direct consequence of the level set formulation.

If (5.4) holds, then necessarily $\rho_{TF} > 0$ on $\{(r,z) \in \partial\Omega : r = 0\}$, and the form (5.1) of $g$ implies that $g \notin L^1(\Omega)$. In this case, the results of Section 2 are still available, and these imply conclusions (1) - (3) of the theorem. Moreover, it is clear that $|u_o|g = \alpha r\rho_{TF} \leq C\rho_{TF} = Cf$, so the hypotheses of Proposition 3.1 (see also Remark 3.2) are satisfied, and it follows that conclusion (4) holds.

We next prove conclusion (6). The inequality $w_\alpha \geq 0$ follows by noticing that $I^\alpha(0 \vee w) \leq I^\alpha(w)$ for all $w \in Dom\, J$. The inequality $w_\alpha \leq \alpha r^2$ is equivalent to the inclusions

$$E_t \subset H_t^\alpha \qquad \text{for a.e. } t > 0, \tag{5.8}$$

where we set $H_t^\alpha = \{\alpha r^2 > t\}$. To prove (5.8) it is sufficient to show that

$$I_t^\alpha(E \cap H_t^\alpha) \leq I_t^\alpha(E) \qquad \text{for admissible any } E \subset \Omega. \tag{5.9}$$

In fact, recalling that $\frac{\partial \rho_{TF}}{\partial r} \leq 0$ and $\alpha r^2 \leq t$ in $E \setminus H_t^\alpha$, we have

$$\int_{\partial^*(E \cap H_t^\alpha)} \rho_{TF} \leq \int_{\partial^* E} \rho_{TF} \qquad \text{and} \qquad \int_{E \setminus H_t^\alpha} \frac{2\rho_{TF}}{r}(\alpha r^2 - t) \leq \int_E \frac{2\rho_{TF}}{r}(\alpha r^2 - t)$$

for any admissible $E$, which gives (5.9).

It follows that the hypotheses of Lemma 4.1 are satisfied (in particular condition (4.2) is a consequence of conclusion (6)), and hence that $w \in BV(\Omega)$.

Next we prove that $\frac{\partial w}{\partial \nu} = 0$ on $\partial\Omega \cap \{r > 0\}$. It suffices to show that for a.e. $t > 0$, the conclusions of Lemma 4.2 hold for $\partial^* E_t$. The lemma as stated does not immediately apply here, since it assumes that $g \in W^{1,\infty}(\Omega)$. But the proof of Lemma 4.2 only requires $g$ to be Lipschitz near $\partial^* E_t$, and this is the case here by (5.8), so the conclusion follows.

Finally, if $\frac{\partial \rho}{\partial r} < 0$ in $\Omega$, then we have $\partial E_t \cap \partial\Omega \neq \emptyset$, since otherwise one could move $E_t$ in the $r$-direction and decrease its energy. Moreover, since $\alpha r^2 - t < 0$ in $E_t$, it also follows that $E_t$ "cannot have holes", that is, each connected component of $E_t$ is simply connected. □

We now illustrate a consequence of conclusion (5.6) above.

**Lemma 5.4.** *Assume the hypotheses of Theorem 5.1 and assume that $\frac{\partial \rho}{\partial r} \leq 0$ in $\Omega$. For $t > 0$, if there exists a 1-form $\zeta^t$ such that*
  (1) $d\zeta^t = -2\rho_{TF}(\alpha r - \frac{t}{r})dr \wedge dz$ *in* $\Omega$
  (2) $\zeta^t = 0$ *on on* $\{(r,z) \in \partial\Omega : r > 0\}$

(3) $\rho_{TF} - |\zeta^t| > 0$ in $H_t^\alpha = \{(r,z) \in \Omega : \alpha r^2 > t\}$

then $\|w_\alpha\|_\infty \leq t$.

*Proof.* Properties of $\zeta^t$ and Stokes Theorem imply that for any nonempty admissible $E$,

$$I_t^\alpha(E) = \int_{\partial^* E} \rho \, d\mathcal{H}^1 + \int_{\partial^* E} \zeta^t \geq \int_{\partial^* E} (\rho - |\zeta^t|) d\mathcal{H}^1 > 0,$$

where $\partial^* E_t$ is oriented in the standard way in the integral $\int_{\partial^* E} \zeta^t$. The final inequality above uses conclusion (6) of Theorem 5.1. It follows that $E_t$, as the minimizer of $I_t^\alpha$, must be the empty set, and hence that $\|w_\alpha\|_\infty \leq t$ by (5.6). □

On the other hand, (5.6) also implies lower bounds for $\|w_\alpha\|_{L^\infty}$, such as

$$\text{if} \quad i_t^\alpha(\gamma) := I_\alpha^t\Big(\Omega \cap \{r > \gamma\}\Big) < 0 \ \text{ for some } \gamma > 0, \qquad \text{then} \quad \|w_\alpha\|_\infty \geq t. \qquad (5.10)$$

**Example 5.5.** We now consider the model case arising from a quadratic trapping potential:

$$\rho_{TF} = \rho_0 - \frac{r^2}{a^2} - \frac{z^2}{b^2}, \qquad \rho_0, a, b \text{ positive constants.} \qquad (5.11)$$

In this case, a 1-form $\zeta$ satisfying the conditions of Lemma 5.4 is given by $\zeta^t = \eta^t(r,z) \, dz$, where

$$\eta^t := \frac{a^2 \alpha}{2} \rho_{TF}^2 - t\left(2R^2 \ln\left(\frac{aR}{r}\right) - \rho_{TF}\right), \qquad \text{where} \quad R = R(z) := \left(\rho_0 - \frac{z^2}{b^2}\right)^{\frac{1}{2}}. \qquad (5.12)$$

Since $\eta^t = 0$ on $\{(r,z) \in \partial\Omega : r > 0\}$, and $\frac{\partial \eta^t}{\partial r} < 0$ in $H_t^\alpha$, we see that $\eta^t > 0$ in $H_t^\alpha$. Thus

$$\|w_\alpha\|_\infty \leq \bar{t}(\alpha) := \inf\left\{t \in [0, \alpha a^2] \,\Big|\, (\rho_{TF} - \eta^t)(r,z) > 0 \ \text{ in } H_t^\alpha\right\}, \qquad (5.13)$$

for $\eta^t$ as in (5.12).

On the other hand, a direct computation shows that the function $i_t^\alpha(\cdot)$ appearing in (5.10) is

$$i_t^\alpha(\gamma) = \frac{4b}{3} Z^3 - \frac{8b}{15} \alpha a^2 Z^5 + t\left[\frac{8b}{3} \rho_0^{3/2} \tanh^{-1}\left(\frac{Z}{\sqrt{\rho_0}}\right) - \frac{8b}{9} Z^3 - \frac{8b}{3} \rho_0 Z\right],$$

where $Z = Z(\gamma) := (\rho_0 - \frac{\gamma^2}{a^2})^{1/2}$. Thus (5.10) becomes

$$\|w_\alpha\|_\infty \geq \underline{t}(\alpha) := \sup\left\{t \in [0, \alpha a^2] \,\Big|\, \min_{\gamma \in (0, a\sqrt{\rho_0})} i_t^\alpha(\gamma) < 0\right\} \qquad (5.14)$$

for $i_\alpha^t$ displayed above.

Both $\bar{t}(\alpha)$ and $\underline{t}(\alpha)$ are easy computed numerically with considerable accuracy. In the next section we report the results of some such computations, and we compare the results to numerical solutions of the minimization problem. These results show that the upper and lower bounds for $\|w_\alpha\|_\infty$ given by (5.13) and (5.14) are often surprisingly close to each other.

**Remark 5.6.** The $t = 0$ functional $I_0^\alpha(\cdot)$ determines whether or not $\|w_\alpha\|_\infty > 0$, and also determines the curve around which vortices first appear, at the critical value for $\alpha$ at which $w_\alpha$ becomes nontrivial. Following Lemma 5.4 and (5.12), we have

$$I_0^\alpha(E) \ = \ \int_{\partial^* E} \rho \, d\mathcal{H}^1 + \frac{1}{2}\alpha a^2 \int_{\partial^* E} \rho^2 \, dz.$$

This functional is considered in [2], where it is shown that if $b > a$ in (5.11), then for every $\alpha > 0$, the minimizer is either $\Omega$ or the empty set, whereas if $a > c_0 \cdot b$, where $c_0$ is an explicit constant, and $\alpha$ is not too large, then $\Omega$ does not minimize $I_0$.

## 6. Numerical results

Since the functionals studied in this paper are weighted versions of the Rudin-Osher-Fatemi denoising model [25], we can compute minimizers by adapting techniques developed in the image processing literature. In the computations reported below, we have followed a modified version of a primal-dual algorithm [26], delineated in Algorithm 1 below, to compute minimizers of the functional $I^\alpha$, from Theorem 5.1, in the model case of a quadratic trapping potential, in which $\rho_{TF}$ is given by (5.11). We recall that the algorithm numerically solves the Euler-Largrange equations (2.20), see also Lemma 2.4, which in the present case may be written formally as as

$$-r \operatorname{div}\left(\rho_{TF} \frac{Du}{|Du|}\right) + 2\rho_{TF}(u - \alpha r^2) = 0. \tag{6.1}$$

These are augmented with boundary conditions from conclusions (6) and (7) of Theorem 5.1. Since the problem is formally overdetermined on $\partial\Omega \cap \{r = 0\}$, we only enforce the Dirichlet condition $u = 0$ there.

---

**Algorithm 1**: Modified Zhu-Chan primal-dual algorithm.

**Data**: Given $\alpha$ and $TOL$ (tolerance for stopping)

Initialize: $k = 0, u^0 = \alpha r^2, \bar{p}^0 = \frac{\nabla u^0}{|\nabla u^0|}$ ;

**while** $\|u^k - u^{k-1}\|_\infty > TOL$ **do**

    Choose time discretization parameters $\tau_1$, and $\tau_2$;
    (we follow [26] and choose $\tau_1 = 0.2 + 0.08k$, $\tau_2 = \frac{0.5 - 5/(15+k)}{\tau_1}$.) ;
    $\bar{p}^{k+1} = \bar{p}^k - \tau_1 \left(\frac{2\rho}{r} \nabla u^k\right)$;
    $\bar{p}^{k+1} = \frac{\bar{p}^{k+1}}{\max\{|\bar{p}^{k+1}|, 1\}}$;
    $u^k = u^{k-1} + \tau_2 \left(2\rho(\alpha r^2 - u^k) + r \, \operatorname{div}(\rho \bar{p}^k)\right)$;
    Set $u^k(r = 0) = 0$;
    $k = k + 1$;

**end**

---

We performed computations for the lower and upper limits of $\|w_\alpha\|_\infty$ found in (5.13) and (5.14) above, and we compared these to $\|\hat{w}_\alpha\|_\infty$ for an approximation $\hat{w}_\alpha$, obtained using Algorithm 1, of the minimizer $w_\alpha$.

In Tables 1 and 2 below we show the results of these computations, for a range of values of the parameter $\alpha$ and for two different choices of the parameters $a$ and $b$ appearing in the definition (5.11) of $\rho_{TF}$. Numerical minimizers $\hat{w}_\alpha$ for several choices of parameters are displayed in Figures 1 and 2. The computations show clearly that as the rotational forcing increases, the size of the vortex-free zone (the "flat spot") decreases but does not vanish.

| $\alpha$ | $\frac{\underline{t}(\alpha)}{\alpha}$ | $\|w_\alpha\|_\infty$ | $\frac{\bar{t}(\alpha)}{\alpha}$ |
|---|---|---|---|
| 5 | 0.140 | 0.146 | 0.21 |
| 10 | 0.321 | 0.320 | 0.39 |
| 20 | 0.485 | 0.490 | 0.54 |
| 40 | 0.6185 | 0.620 | 0.66 |
| 80 | 0.722 | 0.723 | 0.76 |
| 160 | 0.799 | 0.800 | 0.82 |
| 320 | 0.855 | 0.856 | 0.87 |
| 640 | 0.8967 | 0.896 | 0.91 |
| 1280 | 0.926 | 0.925 | 0.93 |
| 2560 | 0.9477 | 0.946 | 0.95 |

TABLE 1. Numerical approximations of $\frac{\underline{t}(\alpha)}{\alpha}$, $\|w_\alpha\|_\infty$, and $\frac{\bar{t}(\alpha)}{\alpha}$ for various values of $\alpha$. The experiments are performed for pancake-shaped regions i.e. $\rho_0 = 1, a = 1$, and $b = 0.5$.

## APPENDIX A. A WEIGHTED RELATIVE ISOPERIMETRIC INEQUALITY

In this Appendix we present the proof of the weighted relative isoperimetric inequality.

*Proof of Proposition 3.4.* **Step 1**. We first note that it suffices to show that there exists some $c_0 > 0$ such that

$$\int_E \rho^m(x)\, dx \leq C \left[\int_{\partial^* E \cap \Omega} \rho^m(x) d\mathcal{H}^{n-1}\right]^{\frac{n+m}{n+m-1}} \qquad \text{if } \int_E \rho^m(x)\, dx \leq c_0. \qquad (A.1)$$

Indeed, for any $0 < c_0 \leq \frac{1}{2}\int_\Omega \rho^m$, if we define

$$c_1(c_0, \Omega) := \inf\left\{\int_{\partial^* E \cap \Omega} \rho^m d\mathcal{H}^{n-1} : c_0 \leq \int_E \rho^m \leq \frac{1}{2}\int_\Omega \rho^m\right\},$$

| $\alpha$ | $\frac{\underline{t}(\alpha)}{\alpha}$ | $\|w_\alpha\|_\infty$ | $\frac{\bar{t}(\alpha)}{\alpha}$ |
|---|---|---|---|
| 20 | 0.0350 | 0.038 | 0.05 |
| 40 | 0.0803 | 0.085 | 0.10 |
| 80 | 0.1211 | 0.125 | 0.14 |
| 160 | 0.1546 | 0.158 | 0.17 |
| 320 | 0.1802 | 0.183 | 0.19 |
| 640 | 0.1997 | 0.201 | 0.21 |
| 1280 | 0.2137 | 0.214 | 0.22 |
| 2560 | 0.2242 | 0.224 | 0.23 |

TABLE 2. Numerical approximations of $\frac{\underline{t}(\alpha)}{\alpha}$, $\|w_\alpha\|_\infty$, and $\frac{\bar{t}(\alpha)}{\alpha}$ for various values of $\alpha$. The experiments are performed for cigar-shaped regions i.e. $\rho_0 = 1, a = 0.5$, and $b = 1$.

then standard arguments imply that the infimum is attained. The fact that $\Omega$ is smooth and connected implies that
$$c_1(c_0, \Omega) > 0,$$
and this fact and (A.1) together imply (3.4).

**Step 2**. For $\delta > 0$ to be fixed below, we will write
$$\Omega_\delta := \{x \in \Omega : \rho(x) < \delta\}, \qquad \Omega^\delta := \{x \in \Omega : \rho(x) > \delta\},$$
We will chose $\delta$ small enough that $\Omega^\delta$ is connected and $\partial \Omega^\delta$ is smooth, and moreover $\Omega_\delta$ consists of a finite number of connected components, say $\Omega_i$ for $i = 1, \ldots, k$, one for each component of $\partial\Omega$. (We will also impose a couple of additional smallness conditions on $\delta$ below, see (A.8), (A.9).) We will write
$$E^\delta = E \cap \Omega^\delta, \qquad E_\delta = E \cap \Omega_\delta, \qquad E_i = E \cap \Omega_i, \quad i = 1, \ldots, k.$$
In view of (A.1), we may assume that
$$\int_E \rho^m \leq \frac{1}{2} \min\left\{ \int_{\Omega_1} \rho^m, \ldots, \int_{\Omega_k} \rho^m, \int_{\Omega^\delta} \rho^m \right\}. \tag{A.2}$$
Since $\partial \Omega^\delta$ is smooth, it follows from (A.2) and the relative isoperimetric inequality in $\Omega^\delta$ (see for example [18] section 4.5.2) that
$$\int_{E^\delta} \rho^m \leq C|E^\delta| \leq C \left[ \mathcal{H}^{n-1}(\partial^* E^\delta \cap \Omega^\delta) \right]^{n/n-1} \leq C \left[ \int_{\partial^* E \cap \Omega^\delta} \rho^m d\mathcal{H}^{n-1} \right]^{n/n-1}$$
where all constants may depend on $\delta$ as well as $n, \Omega, m$. If $\int_{E^\delta} \rho^m > \frac{1}{2}\int_E \rho^m$, then (A.1) follows easily, since then $\int_E \rho^m \leq C(\int_{E^\delta} \rho^m)^\theta$ for $\theta := \frac{n+m}{n+m-1}\frac{n-1}{n} \in (0,1)$. So we may assume that this inequality fails, *i.e.* that
$$\int_{E_\delta} \rho^m \geq \frac{1}{2}\int_E \rho^m. \tag{A.3}$$

Once we assume this, in view of (A.2) it is enough to prove the following weighted relative isoperimetric inequality in $\Omega_i$ for every $i = 1, \ldots, k$:

$$\text{if } \int_{E_i} \rho^m \leq \frac{1}{2} \int_{\Omega_i} \rho^m, \qquad \text{then } \int_{E_i} \rho^m \leq C_i \left( \int_{\partial^* E_i \cap \Omega_i} \rho^m d\mathcal{H}^{n-1} \right)^{\frac{n+m}{n+m-1}}. \tag{A.4}$$

Indeed, once this is know, then for $E$ satisfying (A.2) and (A.3), we have

$$\int_E \rho^m \leq 2 \sum_i \int_{E_i} \rho^m \leq C \sum_i \left( \int_{\partial^* E_i \cap \Omega_i} \rho^m d\mathcal{H}^{n-1} \right)^{\frac{n+m}{n+m-1}}$$

$$\leq C \left( \sum_i \int_{\partial^* E \cap \Omega_i} \rho^m d\mathcal{H}^{n-1} \right)^{\frac{n+m}{n+m-1}} \leq C \left( \int_{\partial^* E \cap \Omega} \rho^m d\mathcal{H}^{n-1} \right)^{\frac{n+m}{n+m-1}}.$$

**Step 3**. We now introduce auxiliary functions that will be used in the proof of (A.4). We will extend $\rho$ to the complement of $\Omega$ by setting

$$\rho(x) := \begin{cases} \text{dist}(x, \partial\Omega) & \text{if } x \in \bar{\Omega} \\ -\text{dist}(x, \partial\Omega) & \text{if not.} \end{cases}$$

Thus the extended $\rho$ is the signed distance to $\partial\Omega$. We recall some standard facts about the signed distance function. First, because $\Omega$ is bounded and $\partial\Omega$ smooth, $\rho$ is smooth in

$$N_\delta := \{x \in \mathbb{R}^n : |\rho(x)| \leq \delta\},$$

for all sufficiently small $\delta > 0$, and in this set,

$$\pi(x) := x - \rho(x)\nabla\rho(x) \text{ is the (unique) closest point in } \partial\Omega \text{ to } x.$$

In addition,

$$\nabla\rho(x) = \nu(\pi(x)), \quad \text{where } \nu(y) = \text{ the inward unit normal to } \partial\Omega \text{ at } y \in \partial\Omega.$$

In particular,

$$|\nabla\rho|^2 = 1, \qquad \text{and by differentiating,} \qquad \nabla^2\rho\,\nabla\rho = 0. \tag{A.5}$$

For $(x, y) \in N_\delta \times \mathbb{R}^m$ we will write

$$\tilde{\rho}(x, y) := (\rho^2(x) + |y|^2)^{1/2} = \text{ the (unsigned) distance from } (x, y) \text{ to } \partial\Omega \times \{0\}$$

$$\tilde{\Omega}_\delta := \{(x, y) : \tilde{\rho}(x, y) < \delta\}.$$

Note also that for $(x, y) \in \tilde{\Omega}_\delta$, the closest point in $\partial\Omega \times \{0\}$ to $(x, y)$ is $\tilde{\pi}(x, y) := (\pi(x), 0)$.

Now we define $F : \tilde{\Omega}_\delta \to \Omega_\delta$ by

$$F(x, y) = \pi(x) + \tilde{\rho}(x, y)\nabla\rho(x) = x + [\tilde{\rho}(x, y) - \rho(x)]\nabla\rho(x).$$

We note for future reference that $F^{-1}(x)$ is a $m$-dimensional sphere of radius $\rho(x)$ for every $x \in \Omega_\delta$; This can be seen by inspecting the definition of $F$, or it can be deduced from the formula

$$F^{-1}(x) = \{(\pi(x) + s\nabla\rho(x), y) : s^2 + |y|^2 = \rho(x)^2\}, \tag{A.6}$$

which is not hard to check.

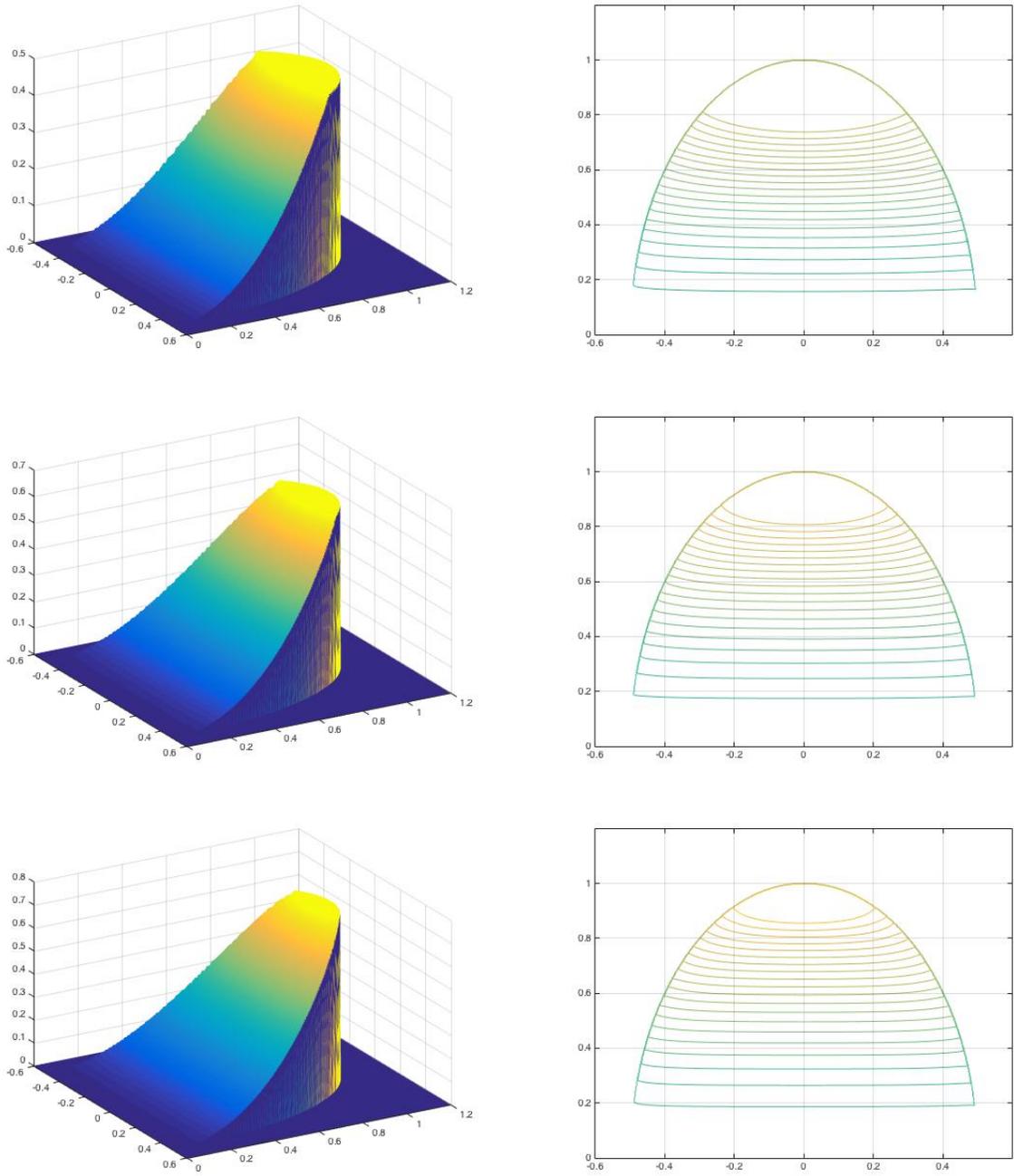

FIGURE 1. minimizer $w_\alpha$ for a pancake-shaped condensate, with parameters as in Table 1, for $\alpha = 20, 40, 80$ from top to bottom.
Left: vertical axis is $w_\alpha/\alpha$. Note the change in vertical scale.
Right: level sets of minimzers.

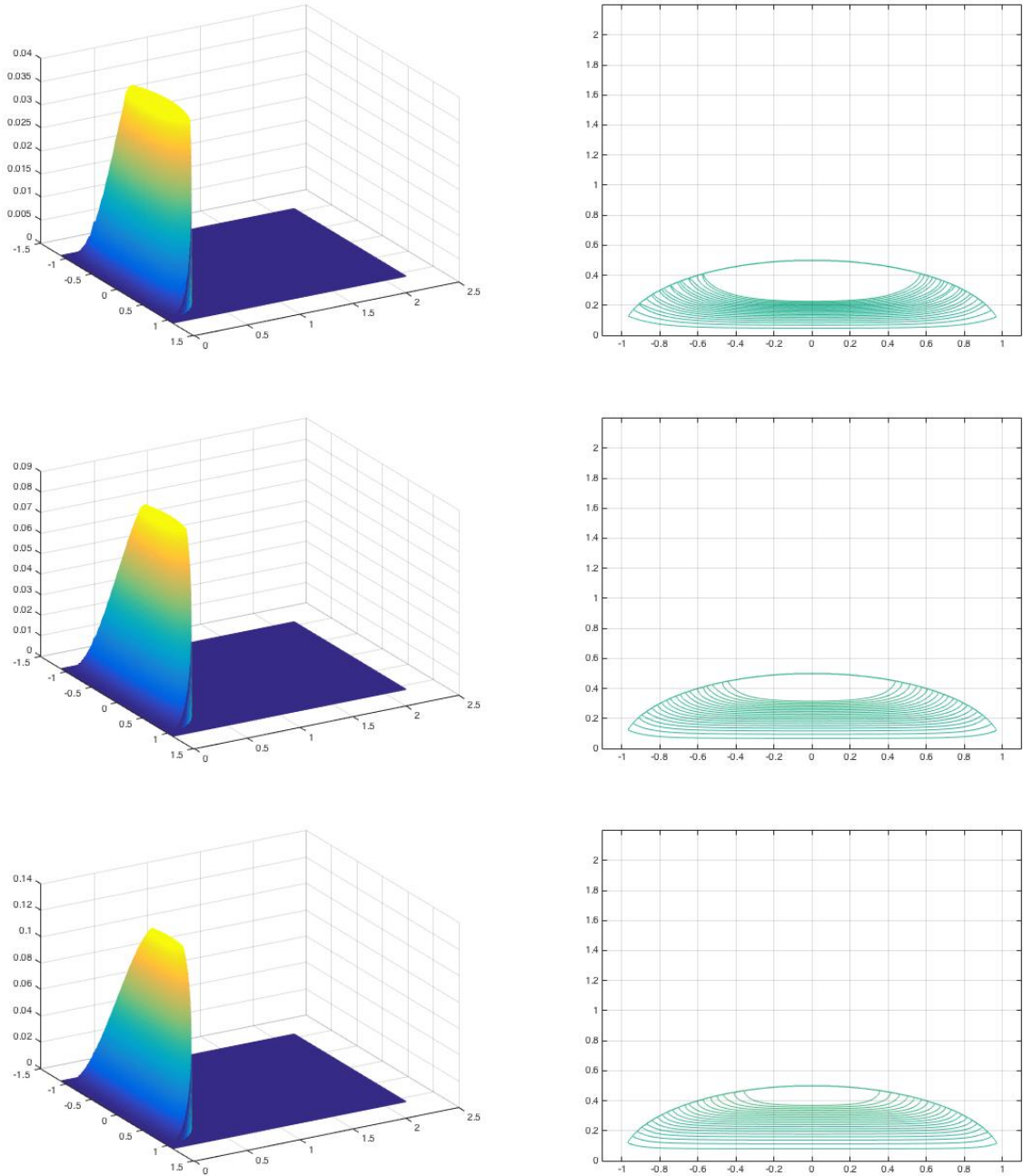

FIGURE 2. minimizer $w_\alpha$ for a cigar-shaped condensate, with parameters as in Table 2, for $\alpha = 20, 40, 80$ from top to bottom.
Left: vertical axis is $w_\alpha/\alpha$. Note the change in vertical scale.
Right: level sets of minimzers.

**Step 4**. It suffices to prove (A.4) for $E_i$ that $\partial E_i \cap \Omega_i$ is smooth, since if we know (A.4) for smooth sets, then the general case follows by a rather standard approximation argument (which the weights do not greatly complicate). It is convenient to write

$$E_i^+ = E_i, \qquad E_i^- = \Omega_i \setminus E_i, \qquad \tilde{E}_i^\pm = F^{-1}(E_i^\pm) \subset F^{-1}(\Omega_i) =: \tilde{\Omega}_i.$$

Note that $\partial \tilde{E}_i$ is smooth in $\tilde{\Omega}_i$ away from $\partial\Omega_i \times \{0\}$, and that $\partial \tilde{E}_i = F^{-1}(\partial E_i)$. Also, $\tilde{\Omega}_i$ has smooth boundary, so the relative isoperimetric inequality guarantees that there is some $C = C(\tilde{\Omega}_i) < \infty$ such that

$$\min_{\pm} \mathcal{H}^{n+m}(\tilde{E}_i^\pm) \leq C \left[ \mathcal{H}^{n+m-1}(\partial \tilde{E}_i \cap \tilde{\Omega}_i) \right]^{\frac{n+m}{n+m-1}}. \tag{A.7}$$

We will show below that, if $\delta$ is small enough, then

$$JF \leq 2 \qquad \text{in } \tilde{\Omega}_\delta, \tag{A.8}$$

where $JF$ denotes the Jacobian of $F$; and that, writing $F_\partial := F|_{\partial \tilde{E}}$

$$\frac{1}{2} \leq J_{n-1} F_\partial \qquad \text{in } \partial \tilde{E}, \tag{A.9}$$

where $J_{n-1} F_\partial$ denotes the $(n-1)$-dimensional Jacobian, whose relevant properties are recalled below. We will first use these two facts to give the proof of (A.4), and then we give the details of (A.8) and (A.9).

With this notation, we use the coarea formula to compute

$$\int_{E_i^+} \rho^m \stackrel{(A.2)}{=} \min_\pm \int_{E_i^\pm} \rho^m(x) dx \stackrel{(A.6)}{=} c(m) \min_\pm \int_{E_i^\pm} \mathcal{H}^m(F^{-1}(x)) dx$$

$$= c(m) \min_\pm \int_{\tilde{E}_i^\pm} JF(x,y)\, dx\, dy \stackrel{(A.8)}{\leq} C \min_\pm \mathcal{H}^{n+m}(\tilde{E}_i^\pm),$$

Similarly, using (more sophisticated) the coarea formula from Federer [18], section 3.2.22,

$$\mathcal{H}^{n+m-1}(\partial \tilde{E}_i \cap \tilde{\Omega}_\delta) \stackrel{(A.9)}{\leq} C \int_{\partial \tilde{E}_i \cap \tilde{\Omega}_\delta} JF_\partial \, d\mathcal{H}^{n-1} = C \int_{\partial E_i \cap \Omega} \rho^m(x) d\mathcal{H}^{n-1}.$$

Combining these two inequalities with (A.7), we deduce that (A.4) holds.

**Step 5**. To complete the proof of the proposition, we verify (A.8) and (A.9). We fix $(x,y) \in \tilde{\Omega}_\delta$ and write $M := DF(x,y)$. Then a computation shows that $M = M_0 + M_1$ where $M_0, M_1$ are $n \times (n+m)$ matrices defined (in block form) by

$$M_0 = (Id_n + (\frac{\rho}{\tilde{\rho}} - 1)\nabla\rho \otimes \nabla\rho, \nabla\rho \otimes \frac{y}{\tilde{\rho}}), \qquad M_1 = (\tilde{\rho} - \rho)(\nabla^2\rho, 0).$$

This means that for $V = (v,w) \in \mathbb{R}^n \times \mathbb{R}^m$ (understood as a column vector),

$$M_0 V = v + \left[ (\frac{\rho}{\tilde{\rho}} - 1)(v \cdot \nabla\rho) + \frac{w \cdot y}{\tilde{\rho}} \right] \nabla\rho. \tag{A.10}$$

One easily checks from the definitions that that $|\tilde{\rho} - \rho| \leq \delta$ in $\tilde{\Omega}_\delta$, so

$$|M_1 V| \leq \delta \|\nabla^2 \rho\|_\infty |v| = C\delta |V|, \tag{A.11}$$

It also follows directly that
$$\ker M_0 = \{(\sigma \nabla \rho(x), w) : \sigma \rho(x) + w \cdot y = 0\},$$
and can also check from (A.10) that
$$M_0 \text{ is a linear isometry of } (\ker M_0)^\perp \text{ onto } \mathbb{R}^n. \tag{A.12}$$
Indeed, if we fix an orthonormal basis $\{v_1, \ldots, v_n\}$ of $\mathbb{R}^n$ such that $v_1 = \nabla \rho(x)$, then the vectors defined by
$$V_1 = \frac{1}{\tilde{\rho}(x,y)}(\rho(x)\nabla \rho(x), y), \qquad V_i = (v_i, 0) \text{ for } i = 2, \ldots, n,$$
form an orthonormal basis for $(\ker M_0)^\perp$ that satisfies $M_0 V_i = v_i$ for all $i$. We now let $V_{n+1}, \ldots, V_{n+m}$ be an orthonormal basis for $\ker M_0$, so that $\{V_i\}_{i=1}^{n+m}$ is an orthonormal basis for $\mathbb{R}^{n+m}$. The Jacobian is
$$JF(x,y) = \left( \sum_{1 \leq i_1 < \ldots < i_n \leq n+m} |MV_{i_1} \wedge \ldots \wedge MV_{i_n}|^2 \right)^{1/2}$$
$$\stackrel{(A.11)}{=} \left( \sum_{1 \leq i_1 < \ldots < i_n \leq n+m} |M_0 V_{i_1} \wedge \ldots \wedge M_0 V_{i_n}|^2 \right)^{1/2} + O(\delta) = 1 + O(\delta).$$
Thus (A.8) follows by a suitable choice of $\delta$. To verify (A.9), we compute $J_{n-1} F_\partial(x,y)$ for some aribitrary $(x,y) \in \partial \tilde{E}$. By definition,
$$J_{n-1} F_\partial(x,y) = \left( \sum_{1 \leq i_1 < \ldots < i_{n-1} \leq n+m-1} |M \tilde{V}_{i_1} \wedge \ldots \wedge M \tilde{V}_{i_{n-1}}|^2 \right)^{1/2}$$
where $\{\tilde{V}_i\}_{i=1}^{n+m-1}$ form an orthonormal basis for $T_{(x,y)} \partial \tilde{E}$. $\mathbb{R}^{n+m}$ which must contain $\ker M_0(x,y)$, since this is just the tangent space at $(x,y)$ to $F^{-1}(F(x,y)) \subset \partial \tilde{E}$. Thus $T_{(x,y)} \partial \tilde{E}$ has the form
$$T_{(x,y)} \partial \tilde{E} = \ker M_0 \oplus L, \qquad \text{for some } n-1\text{-dim. subspace } L \subset \ker M_0^\perp,$$
And according to (A.12), $M_0$ is a linear isometry of $L$ onto its image. So we can choose a basis $\{\tilde{V}_i\}_{i=1}^{n+m-1}$ such that $M_0 \tilde{V}_i = 0$ for $i \geq n$ and $\{M_0 \tilde{V}_i\}_{i=1}^{n-1}$ form an orthonormal set. Then as above we find that $J_{n-1} F_\partial(x,y) = 1 + O(\delta)$, and we conclude by a suitable choice of $\delta$.

$\square$

Department of Mathematics, University of Toronto, Toronto, Ontario, Canada
*E-mail address*: prashant@math.utoronto.ca, rjerrard@math.toronto.edu

Dipartimento di Matematica, Università di Pisa, Pisa, Italy
*E-mail address*: novaga@dm.unipi.it

Dipartimento di Informatica, Università di Verona, Verona, Italy
*E-mail address*: giandomenico.orlandi@univr.it